\documentclass[twocolumn,10pt]{IEEEtran}
\normalsize
\usepackage{bbm}
\usepackage{adjustbox}
\usepackage{enumitem}
\usepackage{cite,algorithm,algorithmic,amsmath,amssymb,amsthm,empheq,mhsetup}
\usepackage{subfigure,amsfonts,balance}
\usepackage{epstopdf}
\usepackage{setspace}
\usepackage[dvipsnames]{xcolor}
\usepackage[left=0.6in, right=0.59in, top=0.6in, bottom=0.59in,centering]{geometry}
\DeclareMathOperator{\EEE}{\mathbb{E}}

\DeclareMathOperator{\aaa}{\pmb{a}}
\DeclareMathOperator{\FF}{\mathcal{F}}

\DeclareMathOperator{\K}{\mathcal{K}}

\DeclareMathOperator{\HHH}{\mathcal{H}}

\DeclareMathOperator{\G}{\pmb{G}}

\DeclareMathOperator{\LL}{\mathcal{L}}

\DeclareMathOperator{\CN}{\mathcal{CN}}

\DeclareMathOperator{\NN}{\mathcal{N}}

\DeclareMathOperator{\e}{\pmb{e}}

\DeclareMathOperator{\rr}{\pmb{r}}
\DeclareMathOperator{\x}{\pmb{x}}
\DeclareMathOperator{\ttt}{\pmb{t}}

\DeclareMathOperator{\y}{\pmb{y}}

\DeclareMathOperator{\Ss}{\pmb{S}}

\DeclareMathOperator{\uu}{\pmb{u}}

\DeclareMathOperator{\g}{\pmb{g}}

\DeclareMathOperator{\ETA}{\pmb{\eta}}

\newtheorem{remark}{Remark}
\newtheorem{proposition}{Proposition}

\ifCLASSINFOpdf
\else
\fi

\begin{document}
	\bstctlcite{IEEEexample:BSTcontrol}
	%
	\title{\Huge Data Size-Aware Downlink Massive MIMO: \\ A Session-Based Approach \vspace{-0mm}}
	%
	%
	\author{Tung~T.~Vu,
		Hien~Quoc~Ngo,
		Minh N. Dao, 
		Michail Matthaiou,
		and Erik G. Larsson
		\thanks{The work of T. T. Vu and H. Q. Ngo was supported by the U.K. Research and Innovation Future Leaders Fellowships under Grant MR/S017666/1. The work of M. N. Dao benefited from the support of the FMJH Program PGMO and EDF. The work of M. Matthaiou was supported by a research grant from the Department for the Economy Northern Ireland under the US-Ireland R\&D Partnership Programme. The work of E. G. Larsson was supported by ELLIIT and the KAW foundation.}
		\thanks{T.~T.~Vu, H.~Q.~Ngo, and M.~Matthaiou are with the Centre for Wireless Innovation (CWI), Queen's University Belfast, Belfast BT3 9DT, UK (e-mail: \{t.vu, hien.ngo,m.matthaiou\}@qub.ac.uk).}
		\thanks{M.~N.~Dao is with the School of Engineering, Information Technology and Physical Sciences, Federation University, VIC 3353, Australia (e-mail: m.dao@federation.edu.au).}
		\thanks{Erik G. Larsson is with Department of Electrical Engineering (ISY), Link\"{o}ping University, SE-581 83 Link\"{o}ping, Sweden (e-mail: erik.g.larsson@liu.se).}
		\vspace{-10mm}
	}
	
		
		\maketitle
		\allowdisplaybreaks
		\begin{spacing}{1}
			\begin{abstract}
				This letter considers the development of transmission strategies for the downlink of massive multiple-input multiple-output networks, with the objective of minimizing the completion time of the transmission.
				Specifically, we introduce a session-based scheme that splits  time into sessions and allocates different rates in different sessions for the different users. 
				In each session, one user is selected to complete its transmission and will not join subsequent sessions, which results in successively lower levels of interference when moving from one session to the next. 
				An algorithm is  developed to assign users and allocate transmit power that minimizes the completion time. Numerical results show that our proposed session-based scheme significantly outperforms  conventional non-session-based schemes.
			\end{abstract}
		\end{spacing}
		
		\vspace{-2mm}

		%
		\IEEEpeerreviewmaketitle
		
		\vspace{-5mm}
		\section{Introduction}
		\vspace{-2mm}
		\label{sec:Introd}
		We are  witnessing an explosion of streaming and learning applications, such as video streaming, live conferencing, and federated learning \cite{nolan21,cicc14TN,vu20TWC}.  
		Many of these applications require computations by  mobile users (UEs)  \cite{yan21TNSM}, and the UEs have fixed amounts of data to receive. 
		To support these applications, it is critical to design transmission schemes that achieve low latency. 
		It is of particular importance to design communication protocols that minimize the completion time, defined as the time it takes for a UE to receive all data destined for it.  
		
		To support the aforementioned applications, massive multiple-input multiple-output (MIMO) can be used due to  its ability to offer high data rates to all UEs simultaneously \cite{ngo16}. 
		To reduce the completion time, conventionally, the achievable rates of all the UEs are maximized via power allocation, and these rates are kept constant during the whole transmission. 
		Another approach is to use different rates during the transmission period \cite{liu17TIT}. 
		When some UEs have already completed their transmissions, other UEs will benefit by having less multi-user interference and higher rates, which results in shorter completion times. 
		In this context, \cite{liu17TIT} studies the completion time for two-user systems from an information-theoretic perspective. 
		However, general schemes for multi-user systems that use different rates within the transmission period have not been explored in the literature. 
		
		\textit{Contributions:} Motivated by \cite{liu17TIT}, we introduce a session-based scheme to reduce the completion times of UEs for the downlink of massive MIMO networks. In this scheme, UE data is transmitted with different rates in different sessions. Specifically, in each session, UEs are assigned so that one UE finishes its transmission and does not participate in subsequent sessions. 
		UEs with uncompleted transmissions are allocated more power to obtain higher achievable rates, and  complete their transmissions faster. 
		Herein, zero-forcing (ZF) processing is used for data transmission. 
		An algorithm is developed for assigning UEs and allocating transmit power, with the objective of  minimizing the completion times of the UEs. 
		Numerical results show that the proposed session-based scheme significantly reduces the completion times compared to conventional transmission that relies on power control only. 
		
		A specific version of this session-based scheme was also used in \cite{vu21ArXiv}, although for a different objective and for a particular application. Herein, we substantially extend \cite{vu21ArXiv} to the general problem of minimizing downlink completion times. If the amounts of UE data in each session are given, then the optimization problem reduces to power and rate control for conventional transmission (with different rate constraints for different users). However, optimizing the amounts of data per session and the thresholds for the rate constraints over multiple sessions is a new and challenging problem.


		
		\vspace{-4mm}
		\section{System Model}\label{sec:SystModel}
		\vspace{-2mm}
		We consider the downlink transmission in a massive MIMO network, where an $M$-antenna base station (BS) serves $K\leq M$ single-antenna UEs simultaneously in the same frequency band.
		Let $S_k$ be the size of the data intended for UE $k$. 
		We  focus on applications where $S_k$ is fixed, such as mobile edge computing and federated learning to name but a few \cite{Feng21,vu20TWC}. 
		We assume that the transmission time is within one large-scale coherence time,\footnote{The large-scale coherence time is the time during which the large-scale fading coefficients remain substantially constant \cite{vu20TWC}.} and the transmission is spans multiple small-scale coherence blocks. 
		A small-scale coherence block is the time-frequency interval over which the channel is substantially static, and is divided into two phases: channel estimation and downlink payload data transmission.
		
		Resource allocation, such as transmit power control, is typically performed to guarantee given quality-of-service targets. We consider two conventional schemes: (i) \textbf{Conventional non-data size-aware scheme}: This scheme allocates power such that all users achieve the same rate  \cite{ngo16}; and (ii) \textbf{Conventional data size-aware scheme}: This scheme allocates power such that each UE receives a rate proportional to the size of its data. This is the traditional scheme used in wireless networks supporting mobile edge computing or federated learning (see, e.g., \cite{Feng21,vu20TWC} and references therein). 
		
		In both conventional schemes, the data rates are kept fixed for the whole transmission. However, since $K$ UEs have different required data sizes, some UEs could complete their transmissions before other UEs. Therefore, optimally, the data rates vary temporally depending on how many active UEs remain in the system.\footnote{A UE that completely receives data from the BS will become inactive.}  The main question is how to allocate power and update the UEs' rates to reduce the completion times. Motivated by this, we next propose the novel session based-scheme.

		\vspace{-4mm}
		\section{Proposed Session-Based Scheme}
\begin{figure}[t!]
			\centering
			\includegraphics[width=0.4\textwidth]{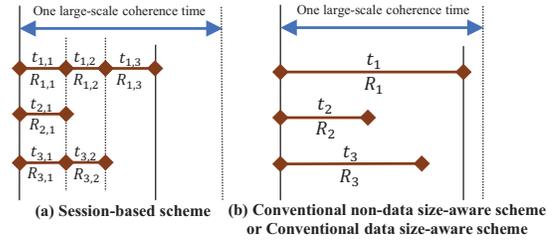}
			\vspace{-4mm}
			\caption{One transmission period with three UEs.}
			\label{fig:completiontime}
		\end{figure}

		\vspace{-1mm}
		Our proposed scheme uses different rates in different time periods for data transmissions.
		The transmission period  during which the rates are kept fixed is called a ``session''.
		More precisely, session $i$ is defined as follows: (i) in session $i$, there are $K-i+1$ active UEs; and (ii) at the beginning of session $i$, the BS updates the rates for these active UEs. These rates will be kept fixed until the end of this session, where one UE completes receiving data from the BS. Thus, the BS transmits data to all $K$ UEs during $K$ sessions. Fig.~\ref{fig:completiontime} illustrates the proposed session-based transmission as well as conventional transmission for a system with three UEs.
		Denote by $a_i$ the indicator that is defined as
		\begin{align}\label{a}
			a_{k,i} \triangleq
			\begin{cases}
				1,& \text{if UE $k$ is receiving data in session $i$,}\\
				0, & \mbox{otherwise}.
			\end{cases}
		\end{align}
		Denote by $\K_i\triangleq\{k|a_{k,i} = 1\}$ the set of $K_i=\sum_{k\in\K}a_{k,i}$ UEs assigned in session $i\in\K$, where $\K=\{1, 2, \ldots, K\}$. Then, we have
		\begin{align}
			\label{sumaki}
			& a_{k,1} = 1, \sum_{k\in\K} a_{k,i} = K-i+1, a_{k,i} \leq a_{k,i-1},  \forall i,
		\end{align}
		to ensure that all the UEs will be served in session $1$. In each subsequent session, one UE completes its transmission and will not join the next sessions. As such, more power is allocated to the UEs that have not yet completed their transmissions. 
		Note that, the conventional (non-session-based) schemes are special cases of the proposed scheme when all UEs are served in session $1$ with $a_{k,1}=1,\forall k$.
		
		\textbf{Uplink channel estimation}:
		In each small-scale coherence block of length $\tau_c$, each UE sends its pilot of length $\tau_{p}$ to the BS. We assume that all pilots are mutually  orthogonal, which requires $\tau_{p}\geq K$.\footnote{One can let only the participating UEs send their pilots in session $i$, i.e., $\tau_{p} = K_i$ to increase the small-scale coherence block length for payload data transmission. However, since $\tau_c$ is normally much larger than $K\geq K_i$ in many applications \cite{vu20TWC}, letting all the UEs send their pilots, i.e., taking $\tau_{p} = K$, has a negligible impact on data rates. On the other hand, the channel estimation is better when the pilot length $\tau_{p} = K>K_i$, which potentially improves the data rates.} Denote by $\g_{k} \!=\! (\beta_{k})^{1/2}\tilde{\g}_{k}$ the channel vector from UE $k$ to the BS, where $\beta_{k}$ and $\tilde{\g}_{k}$ 
		are the large-scale fading coefficient and small-scale fading  vector, respectively.
		With minimum mean-square error (MMSE) estimation, the channel estimate $\hat{\g}_{k}$ of $\g_{k}$ is
		distributed according to $\CN(\pmb{0},\sigma_{k}^2\pmb{I}_M)$, where $\sigma_k^2 = \frac{\tau_{p} \rho_{p} \beta_k^2 }{ \tau_{p} \rho_{p} \beta_k +1 }$, and $\rho_{p}$ is the normalized transmit power of each pilot symbol \cite[(3.8)]{ngo16}.
		Let $\hat{\G}_i\triangleq [\hat{\g}_{1},\dots,\hat{\g}_{K-i+1}], \forall k\in\K_i$, be the matrix stacking the estimated channels of all participated UEs in session $i$.

		\textbf{Downlink payload data transmission}: In session $i$, the BS uses ZF to transmit data to $K-i+1$ UEs. 
		With ZF, the signal transmitted by the BS is given by $\x_{i}\!\!=\!\! \sqrt{\rho}\sum_{k\in \K_i}\uu_{k,i} s_{k,i}$,
		where $\rho$ is the  normalized transmit power at the BS; $s_{k,i}$, with $\EEE\{|s_{k,i}|^2\}=1$, is the symbol intended for UE $k$; and $\uu_{k,i}\! =\! \sqrt{\eta_{k,i}{\sigma}_{k}^2(M\!-\!K_i)} \hat{\G}_i(\hat{\G}_i^H\hat{\G}_i)^{-1}\e_{k,K_i}$ is the ZF precoding vector. Here, $\EEE\{x\}$ denotes the expected value of a random variable $x$, and $\pmb{X}^H$ represents the conjugate transpose of a matrix $\pmb{X}$. In the precoding vector, $\eta_{k,i}$ is a power control coefficient, and $\e_{k,K_i}$ is the $k$-th column of $\pmb{I}_{K_i}$. 
		The transmitted power at the BS is constrained by $\EEE\{|\x_{i}|^2\}\leq \rho$ which is equivalent to
		\begin{align}
			\label{powerdupperbound}
			\sum_{k\in\K_i}\eta_{k,i} \leq 1, \forall i\in\K.
		\end{align}
		We enforce 
		\begin{align}
			\label{eta-a-relation} 
			(\eta_{k,i} = 0, \text{if $a_{k,i} = 0$}), \forall k,i
		\end{align}
		to ensure that the BS will not allocate any power to UEs that are not served in session $i$.
		The achievable rate at UE $k$ in session $i$ is given by \cite[Eq. (3.56)]{ngo16}: $R_{k,i}(\ETA_i) = \frac{\tau_c - \tau_{p}}{\tau_c}B\log_2 \big( 1 + \text{SINR}_{k,i}(\ETA_i)\big)$,
		where $B$ is the bandwidth, 
		$\text{SINR}_{k,i}(\ETA_i) 
		=\frac{(M-K_i)\rho \hat{\sigma}_k^2\eta_{k,i}}
		{\rho (\beta_k - {\sigma}_k^2) \sum_{\ell\in\K_i} \eta_{\ell,i} +1} \overset{\eqref{eta-a-relation}}{=} \frac{(M-K_i)\rho {\sigma}_k^2\eta_{k,i}}
		{\rho (\beta_k - \hat{\sigma}_k^2) \sum_{\ell\in\K} \eta_{\ell,i} +1}$
		is the effective downlink signal-to-interference-plus-noise ratio (SINR), and $\ETA_i\triangleq\{\eta_{k,i}\}_{k\in\K}$.
		
		\textbf{Completion time}:
		Let $S_{k,i}$ be the size of the data sent to UE $k$ in session $i$. Then, we have
		\begin{align}
			\label{samesizedatad}
			& \sum_{i\in\K} S_{k,i} = S_k, \forall k.
		\end{align}
		Let $t_{i}$ be the time duration of session $i$. Then, the transmission time $t_{k,i}$ of UE $k\in\K$ in session $i$  is given by
		\begin{align}\label{tdki}
			t_{k,i} \!=\! a_{k,i} t_{i}, \forall k,i.
		\end{align}
		Thus, 
		\begin{align}
			\label{sametime:sessisionS1}
			S_{k,i} \!=\! R_{k,i}(\ETA_i) t_{k,i} \!\overset{\eqref{tdki}}{=}\! R_{k,i}(\ETA_i) a_{k,i} t_{i} \!\overset{\eqref{eta-a-relation} }{=}\! R_{k,i}(\ETA_i) t_{i}, \forall k,i.
		\end{align}
		Clearly, \eqref{sametime:sessisionS1} also implies that $(S_{d,k,i} = 0, \text{if $a_{k,i} = 0$}), \forall k,i$. The completion time of UE $k$ is the sum of its transmission times across all sessions, i.e., $\sum_{i\in\K}a_{k,i}t_{i}$. 
		Let $\widetilde{T}_c$ and $T_c$ the large-scale and small-scale coherence times, respectively. Since each session spans multiple small-scale coherence blocks but always fits within one large-scale coherence time, we have
		\begin{align}
			\label{small-scale:session:relation}
			T_c\leq t_i, \forall i
			\\
			\label{large-scale:session:relation}
			\sum_{i\in\K} t_i \leq \widetilde{T}_c.
		\end{align}
		\vspace{-2mm}
		\begin{remark}
			In this work, in order to  focus on fundamental principles of the proposed scheme, that is, UE assignment and rate allocation, we consider independent Rayleigh channels. The optimization and analysis of a session-based scheme tailored to correlated channels is interesting, but analytically challenging  and beyond the scope of the paper.
			Thus, such designs are left for future work. 
		\end{remark}


		\vspace{-5mm}
		\section{Completion Time Minimization}
		\label{sec:PF}
		\vspace{-1mm}
		We address the minimization of the completion time of our proposed session-based scheme, and specifically minimizing the longest completion time among the UEs. For comparison, we also include the completion time minimization for the conventional schemes. 
		
		\vspace{-5mm}
		\subsection{Proposed Session-Based Scheme}
		\vspace{-1mm}
		\label{sec:alg:sb}
		The problem of minimizing the completion time of UEs by optimizing the UE assignment ($\aaa$) and transmit power ($\ETA$) in the session-based design is
		\begin{subequations}\label{Pmain:sb}
			\begin{align}
				\label{CFsb}
				\!\!\!\!\!\underset{\x}{\min} \,\,
				& 
				\max_{k} \sum_{i\in\K} a_{k,i}t_i,
				\\
				\!\!\!\!\!\mathrm{s.t.}\,\,
				\nonumber
				& \eqref{a}-\eqref{eta-a-relation},  
				\eqref{samesizedatad}, 
				\eqref{sametime:sessisionS1}-\eqref{large-scale:session:relation}
				\\
				\label{powerlowerbound}
				& 0\leq \eta_{k,i}, \forall k,i,
			\end{align}
		\end{subequations}
		where $\x \triangleq \{\aaa, \ETA,\Ss, \ttt\}$, $\Ss\! \triangleq \{S_{k,i}\}$, $\aaa \triangleq \{a_{k,i}\}$, $\ttt\triangleq\{t_i\}$ $\forall k,i$. 
		
		Finding a globally optimal solution to problem \eqref{Pmain:sb} is challenging due to the mixed-integer and nonconvex constraints \eqref{a}, \eqref{eta-a-relation}, and \eqref{sametime:sessisionS1}. Thus, we instead propose an approach that is suitable for practical implementation. First, we replace constraint \eqref{eta-a-relation} by
		\begin{align}
			\label{power-a-relation}
			\eta_{k,i} \leq a_{k,i}, \forall k,i,
		\end{align}
		and constraint \eqref{sametime:sessisionS1} by
		\begin{align}
			\label{sametime:sessisionS1-1}
			& S_{k,i} \leq R_{k,i}(\ETA_i) a_{k,i} t_{i}, \forall k,i
			\\
			\label{sametime:sessisionS1-2}
			& S_{k,i} \geq R_{k,i}(\ETA_i) a_{k,i} t_{i}, \forall k,i.
		\end{align}
		We further replace constraints \eqref{sametime:sessisionS1-1} and \eqref{sametime:sessisionS1-2} by
		\begin{align}
			\label{ratedki-lowerbound-1}
			& \hat{r}_{k,i} \leq R_{k,i}(\ETA_i), \forall k,i
			\\
			\label{ratedki-upperbound-1}
			& \tilde{r}_{k,i} \geq R_{k,i}(\ETA_i), \forall k,i
			\\
			\label{tdki-lowerbound}
			& \hat{t}_{k,i} \leq a_{k,i} t_{i}, \forall k,i
			\\
			\label{tdki-upperbound}
			& \tilde{t}_{k,i} \geq a_{k,i} t_{i}, \forall k,i
			\\
			\label{sametime:sessisionS1-1-a}
			& S_{k,i} \leq \hat{r}_{k,i} \hat{t}_{k,i}, \forall k,i
			\\
			\label{sametime:sessisionS1-2-a}
			& S_{k,i} \geq \tilde{r}_{k,i} \tilde{t}_{k,i}, \forall k,i,
		\end{align}
		where $\hat{\rr} \triangleq \{\hat{r}_{k,i}\}, \tilde{\rr} \triangleq \{\tilde{r}_{k,i}\}, \hat{\ttt} \triangleq \{\hat{t}_{k,i}\}, \tilde{\ttt} \triangleq \{\tilde{t}_{k,i}\}$ are additional variables. 
		We observe from \eqref{ratedki-lowerbound-1}--\eqref{sametime:sessisionS1-1-a} that $S_{k,i} \leq \tilde{r}_{k,i} \tilde{t}_{k,i}, \forall k,i$.
		Thus, \eqref{sametime:sessisionS1-2-a} is equivalent to
		\begin{align}
			\label{V1}
			V_1(\tilde{\rr}, \tilde{\ttt}, \Ss) &\triangleq \sum_{k\in\K}\sum_{i\in\K} (\tilde{r}_{k,i} \tilde{t}_{k,i} - S_{k,i}) \leq 0.
		\end{align}
		
		Now, in order to handle the binary constraint \eqref{a}, we note that $x\in\{0,1\}\Leftrightarrow x\in[0,1]\,\&\,x-x^2\leq0$ \cite{vu18TCOM}. Thus, \eqref{a} can be replaced by the following equivalent constraints:
		\begin{align}
			\label{V2}
			&\!\!\! V_2(\aaa) \triangleq \sum_{k\in\NN} \sum_{i\in\K} (a_{k,i}\!-\!a_{k,i}^2) \leq 0
			\\
			\label{arelax}
			& \!\! 0\leq a_{k,i}\! \leq 1, \forall k,i.
		\end{align}
		Then, problem \eqref{Pmain:sb} is written into a more tractable form as 
		\begin{subequations}\label{Pmain:epi}
			\begin{align}
				\label{CFPmulti}
				\underset{\widetilde{\x}}{\min} \,\,
				& q
				\\
				\mathrm{s.t.}\,\,
				\nonumber
				& 
				\eqref{sumaki}, 
				\eqref{powerdupperbound},
				\eqref{samesizedatad},
				\eqref{small-scale:session:relation},
				\eqref{large-scale:session:relation},
				\eqref{powerlowerbound},
				\eqref{power-a-relation},
				\eqref{ratedki-lowerbound-1}-\eqref{sametime:sessisionS1-1-a},
				\eqref{V1}-\eqref{arelax}
				\\
				\label{q:lb}
				& \sum_{i\in\K} \tilde{t}_{k,i} \leq q, \forall k,i,
			\end{align}
		\end{subequations}
		where $\widetilde{\x}\!\triangleq\! \{\x, \hat{\rr}, \tilde{\rr}, \hat{\ttt}, \tilde{\ttt}, q\}$, and $q$ is an additional variable. Let $\FF \!\triangleq\!  \{\eqref{sumaki}, 
		\eqref{powerdupperbound},
		\eqref{samesizedatad},
		\eqref{small-scale:session:relation},
		\eqref{large-scale:session:relation},
		\eqref{powerlowerbound},
		\eqref{power-a-relation},
		\eqref{ratedki-lowerbound-1}-\eqref{sametime:sessisionS1-1-a},
		\eqref{V1}-\eqref{arelax}, \eqref{q:lb}\}$ be the feasible set of problem \eqref{Pmain:epi}.
		We consider the problem
		\begin{align}\label{Pmain:epi:relax}
			\underset{\widetilde{\x} \in \widehat{\FF}}{\min} \,\,
			&\LL(\aaa, \tilde{\rr}, \tilde{\ttt}, \Ss, \lambda),
		\end{align}
		where $\LL(\aaa, \tilde{\rr}, \tilde{\ttt}, \Ss, \lambda) \triangleq q\!+\! \lambda( \gamma_1V_1(\tilde{\rr}, \tilde{\ttt}, \Ss) + \gamma_2 V_2(\aaa))$ is the Lagrangian of \eqref{Pmain:epi}, $\gamma_1, \gamma_2 > 0$ are fixed weights, and $\lambda$ is the Lagrangian multiplier corresponding to constraints \eqref{V1}, \eqref{V2}. Here, $\widehat{\FF} \triangleq \FF \setminus \{\eqref{V1}, \eqref{V2}\}$.
		\vspace{-1mm}
		\begin{proposition}
			\label{proposition-dual}
			The values $V_{1,\lambda}$ and $V_{2,\lambda}$ of $V_1$ and $V_2$ at the solution of \eqref{Pmain:epi:relax} corresponding to $\lambda$  converge to $0$ as $\lambda \rightarrow +\infty$. Moreover, problem \eqref{Pmain:epi} has strong duality, i.e.,
			\begin{equation}\label{Strong:Dualitly:hold}
				\underset{\widetilde{\x}\in\FF}{\min}\,\, q
				= \underset{\lambda\geq0}{\sup}\,\,
				\underset{\widetilde{\x}\in\widehat{\FF}}{\min}\,\,
				\LL(\aaa, \tilde{\rr}, \tilde{\ttt}, \Ss, \lambda),
			\end{equation}
			and consequently, \eqref{Pmain:epi} is equivalent to \eqref{Pmain:epi:relax} at the optimal solution $\lambda^* \geq0$ of the sup-min problem in \eqref{Strong:Dualitly:hold}.
		\end{proposition}
		\vspace{-5mm}
		\begin{proof}
			See Appendix.
		\end{proof}
		\vspace{-3mm}
		Theoretically, it is required to have $V_{1,\lambda}=0$ and $V_{2,\lambda}=0$ in order to obtain the optimal solution to problem \eqref{Pmain:epi}. By Proposition~\ref{proposition-dual}, $V_{1,\lambda}$ and $V_{2,\lambda}$ converge to $0$ as $\lambda\to+\infty$. In practice,  it is sufficient to accept $V_{1,\lambda}\leq\varepsilon, V_{2,\lambda}\leq\varepsilon$ for some small $\varepsilon$ with a sufficiently large value of $\lambda$.
		In our numerical experiments, for $\varepsilon = 10^{-3}$, we see that $\lambda=1$ with $\gamma_1=0.1, \gamma_2=0.01$ is enough to ensure that $V_{1,\lambda}\leq\varepsilon, V_{2,\lambda}\leq\varepsilon$. This way of choosing $\lambda$ has been widely used in the literature, e.g.,
		see \cite{vu18TCOM} and references therein.
		
		Problem \eqref{Pmain:epi:relax} is still difficult to solve due to the nonconvex constraints \eqref{ratedki-lowerbound-1}--\eqref{sametime:sessisionS1-1-a}, and nonconvex parts $V_1(\aaa), V_2(\tilde{\rr}, \tilde{\ttt}, \Ss)$ in the cost function $\LL(\aaa, \tilde{\rr}, \tilde{\ttt}, \Ss,\lambda)$. To deal with \eqref{ratedki-lowerbound-1}, we observe that $\log\big(1+\frac{x}{y}\big) \geq \log\big(1+\frac{x^{(n)}}{y^{(n)}}\big) + \frac{2x^{(n)}}{(x^{(n)}+y^{(n)})} 
		- \frac{(x^{(n)})^2}{(x^{(n)}+y^{(n)})x}
		- \frac{x^{(n)}y}{(x^{(n)}+y^{(n)})y^{(n)}}$, 
		where $x > 0, y > 0$ \cite[Eq. (76)]{long21TCOM}. 
		Therefore, the concave lower bound
		$\widehat{R}_{d,k,i} (\ETA_i)$ of $R_{d,k,i}(\ETA_i)$ is given by $\widehat{R}_{d,k,i} \triangleq \frac{\tau_c - \tau_{p}}{\tau_c\log 2} B \Big[ \log\big(1+\frac{\Upsilon_i^{(n)}}{\Phi_i^{(n)}}\big) + \frac{2\Upsilon_i^{(n)}}{(\Upsilon_i^{(n)}+\Phi_i^{(n)})} 
		- \frac{(\Upsilon_i^{(n)})^2}{(\Upsilon_i^{(n)}+\Phi_i^{(n)})\Upsilon_i}
		- \frac{\Upsilon_i^{(n)}\Phi}{(\Upsilon_i^{(n)}+\Phi_i^{(n)})\Phi_i^{(n)}}\Big]$,
		where $\Upsilon_i(\eta_{k,i}) \triangleq (M-K_i)\rho \hat{\sigma}_k^2\eta_{k,i}$ and $\Phi_i(\ETA_i) \triangleq \rho (\beta_k - \sigma_k^2) \sum_{\ell\in\K} \eta_{\ell,i} +1$. Then \eqref{ratedki-lowerbound-1} can be approximated by the following convex constraint
		\begin{align}
			\label{ratedki-lowerbound-1-a}
			& \hat{r}_{k,i} \leq \widehat{R}_{k,i}(\ETA_i), \forall k,i.
		\end{align}
		
		To deal with constraints \eqref{ratedki-upperbound-1}, we observe that
		$\log\big(1+\frac{x}{y}\big) \leq \log\big(x^{(n)}+y^{(n)}\big) + \frac{x + y - x^{(n)} - y^{(n)}}{x^{(n)}+y^{(n)}} - \log(y)$,
		where $x > 0,y > 0$.
		Therefore, the convex upper bound $\widetilde{R}_{k,i}(\ETA_i)$ of 
		$R_{k,i}(\ETA_i)$ is expressed as $\widetilde{R}_{d,k,i} \triangleq \frac{\tau_c - \tau_{u,p}}{\tau_c\log 2} B \Big[\log\big(\Upsilon_i^{(n)}+\Phi_i^{(n)}\big)
		+ \frac{\Upsilon_i + \Phi_i - \Upsilon_i^{(n)} - \Phi_i^{(n)}}{\Upsilon_i^{(n)}+\Phi_i^{(n)}} - \log(\Phi_i)\Big]$.
		Thus, constraint \eqref{ratedki-upperbound-1} can be approximated by the following convex constraint
		\begin{align}
			\label{ratedki-upperbound-1-a}
			\tilde{r}_{k,i} \geq \widetilde{R}_{k,i}(\ETA_i), \forall k,i.
		\end{align}
		
		Next, we observe that 
		$xy \leq 0.25 [(x+y)^2-2(x^{(n)}-y^{(n)})(x-y) + (x^{(n)}-y^{(n)})^2]$ and $-xy \leq 0.25 [(x\!-\!y)^2\!-\!2(x^{(n)}\!+\!y^{(n)})(x\!+\!y)
		+ (x^{(n)}+y^{(n)})^2], \forall x\geq0, y\geq0, z\geq0$ \cite{vu20TWC}. 
		Therefore, \eqref{tdki-lowerbound}--\eqref{sametime:sessisionS1-1-a} can be approximated by the following convex constraints
		\begin{align}
			\nonumber
			\label{tdki-lowerbound-1}
			& \hat{t}_{d,k,i} +0.25 [(a_{k,i}-t_{d,i})^2 -2(a_{k,i}^{(n)}+t_{d,i}^{(n)})(a_{k,i}+t_{d,i})
			\\
			&\qquad\qquad\qquad\qquad\qquad
			+ (a_{k,i}^{(n)}+t_{d,i}^{(n)})^2] \leq 0, \forall k,i
			\\
			\nonumber
			&  0.25[(a_{k,i}+t_{d,i})^2-2(a_{k,i}^{(n)}-t_{d,i}^{(n)})(a_{k,i}-t_{d,i})
			\\
			&\qquad\qquad\qquad\quad + (a_{k,i}^{(n)}-t_{d,i}^{(n)})^2] - \tilde{t}_{d,k,i} \leq 0, \forall k,i
			\\
			\label{sametime:sessisionS1-1-a:cvx}
			\nonumber
			& S_{d,k,i}\! +\! 0.25 [(\hat{r}_{d,k,i}\!-\!\hat{t}_{d,k,i})^2 \!-\! 2(\hat{r}_{d,k,i}^{(n)}\!+\!\hat{t}_{d,k,i}^{(n)})(\hat{r}_{d,k,i}\!+\!\hat{t}_{d,k,i})
			\\
			&\qquad\qquad\qquad\qquad\quad + (\hat{r}_{d,k,i}^{(n)}+\hat{t}_{d,k,i}^{(n)})^2] \leq 0, \forall k,i.
		\end{align}
		Similarly, the convex upper bounds $\widetilde{V}_1(\aaa), \widetilde{V}_2(\tilde{\rr}, \tilde{\ttt}, \Ss)$ of the nonconvex parts $V_1(\aaa), V_2(\tilde{\rr}, \tilde{\ttt}, \Ss)$ are respectively given by
		\begin{align}
			\nonumber
			& \widetilde{V}_1(\tilde{\rr}, \tilde{\ttt}, \Ss)
			\! \triangleq\!\! \sum_{i\in\K}\sum_{k\in\NN}\!
			0.25 [(\tilde{r}_{d,k,i}\!+\!\tilde{t}_{d,k,i})^2\!
			\\
			\nonumber
			&-\!2(\tilde{r}_{d,k,i}^{(n)}\!-\!\tilde{t}_{d,k,i}^{(n)})(\tilde{r}_{d,k,i}\!-\!\tilde{t}_{d,k,i})
			\!+\! (\tilde{r}_{d,k,i}^{(n)}\!-\!\tilde{t}_{d,k,i}^{(n)})^2 \!-\! 4S_{d,k,i}] 
			\\
			\nonumber
			& \widetilde{V}_2(\aaa)
			\!\triangleq\! \sum_{i\in\K}\sum_{k\in\NN} (a_{k,i}-2a_{k,i}^{(n)}a_{k,i} + (a_{k,i}^{(n)})^2).
		\end{align}
		
		\begin{algorithm}[!t]
			\caption{Solving problem \eqref{Pmain:epi:relax}}
			\begin{algorithmic}[1]
				\label{alg}
				\STATE \textbf{Initialize}: Set $n\!=\!0$ and choose a random point $\widetilde{\x}^{(0)}\!\in\!\widehat{\FF}$.
				\REPEAT
				\STATE Update $n=n+1$
				\STATE Solve \eqref{Pmain:epi-approx} to obtain its optimal solution $\widetilde{\x}^*$
				\STATE Update $\widetilde{\x}^{(n)}=\widetilde{\x}^*$
				\UNTIL{convergence}
			\end{algorithmic}
		\end{algorithm}
		
		At iteration $(n+1)$, for a given point $\widetilde{\x}^{(n)}$, problem \eqref{Pmain:epi:relax} can finally be approximated by the following convex problem
		\begin{align}
			\label{Pmain:epi-approx}
			\underset{\widetilde{\x}\in\widetilde{\FF}}{\min} \,\,
			& \widehat{\LL}(\aaa, \tilde{\rr}, \tilde{\ttt}, \Ss,\lambda) 
		\end{align}
		where $\widehat{\LL}(\aaa, \tilde{\rr}, \tilde{\ttt}, \Ss,\lambda) \triangleq q \!+\! \lambda( \gamma_1 \widetilde{V}_1(\aaa) + \gamma_2 \widetilde{V}_2(\tilde{\rr}, \tilde{\ttt}, \Ss)$ and 
		$\widetilde{\FF}\triangleq\{
		\{\eqref{sumaki}, 
		\eqref{powerdupperbound},
		\eqref{samesizedatad},
		\eqref{powerlowerbound},
		\eqref{power-a-relation},
		\eqref{arelax}, \eqref{q:lb},
		\eqref{ratedki-lowerbound-1-a},
		\eqref{ratedki-upperbound-1-a}-\eqref{sametime:sessisionS1-1-a:cvx}\}
		$ is a convex feasible set.
		In Alg.~\ref{alg}, we outline the main steps to solve problem \eqref{Pmain:epi:relax}.
		Starting from a random point $\widetilde{\x}\in\widehat{\FF}$, we solve \eqref{Pmain:epi-approx} to obtain its optimal solution $\widetilde{\x}^*$, and use $\widetilde{\x}^*$ as an initial point in the next iteration. The algorithm terminates when an accuracy level of $\varepsilon$ is reached. 
		Alg.~\ref{alg} will converge to 
		a Fritz John solution of problem \eqref{Pmain:epi:relax} (hence \eqref{Pmain:epi} or \eqref{Pmain:sb}). The proof of this fact is rather standard, and follows from \cite[Proposition 2]{vu18TCOM}. 
		
		Note that Problem \eqref{Pmain:sb} is constructed using the achievable rates (5) that depend only on the large-scale coefficients $\beta_{k}$. Before the downlink transmission, the BS solves  (10) to obtain the session durations, per-session user assignments, data rates, and transmit powers. Therefore, no extra signalling overhead to schedule UEs/rates and no optimization algorithm are required during the transmission.

		\vspace{-4mm}
		\subsection{Conventional Schemes}
		\vspace{-1mm}
		The conventional schemes can be considered as special cases of the session-based scheme with only one session. Thus, all variables in the conventional schemes can be directly obtained from the session-based scheme by dropping the index $i$. More precisely, the power constraint at the BS and the achievable rate of UE $k$ in the conventional schemes are, respectively, given by
		\begin{align}
			\label{powerdupperbound:asyn:syn}
			&\sum_{k\in\K}\eta_{k} \leq 1
		\end{align}
		and $R_{k}(\ETA) = \frac{\tau_c - \tau_{p}}{\tau_c}B\log_2 \big( 1 + \text{SINR}_{k}(\ETA)\big)$, where $\ETA\triangleq\{\eta_{k}\}_{k\in\K}$ are power control  coefficients, and $\text{SINR}_{k}(\ETA) =
		\frac{(M-K)\rho \sigma_k^2\eta_{k}}
		{\rho (\beta_k - \sigma_k^2) \sum_{\ell\in\K} \eta_{\ell} +1}$. 
		Note that since $T_c$ is in order of milliseconds, and the conventional schemes have only one session, the completion times of UEs is normally larger than $T_c$, which is confirmed in the numerical results in Section V.
		
		\subsubsection{Conventional Data Size-Aware Scheme}
		
		The corresponding problem of completion time minimization is
		\begin{subequations}\label{Pmain:conven}
			\begin{align}
				\!\!\!\!\!\underset{\ETA}{\min} \,\,
				& 
				\max_{k} \tfrac{S_{k}}{{R_{k}(\ETA)}}
				\\
				\!\!\!\!\!\mathrm{s.t.}\,\,
				\nonumber
				& \eqref{powerdupperbound:asyn:syn}
				\\
				\label{powerlowerbound:as}
				& 0\leq \eta_{k}, \forall k,
				\\
				& S_k/R_k(\ETA) \leq \widetilde{T}_c.
			\end{align}
		\end{subequations}
		Problem \eqref{Pmain:conven} can be transformed into  epigraph form as
		\begin{subequations}\label{Pmain:conven:epi}
			\begin{align}
				\!\!\!\!\!\underset{\y}{\min} \,\,
				& z
				\\
				\!\!\!\!\!\mathrm{s.t.}\,\,
				\nonumber
				& \eqref{powerdupperbound:asyn:syn}, \eqref{powerlowerbound:as}
				\\
				\label{z:lb}
				& 
				S_k/r_k \leq z \leq \widetilde{T}_c, \forall k
				\\
				\label{Rk:lb}
				& r_k \leq R_k(\ETA), \forall k,
			\end{align}
		\end{subequations}
		where $\y\triangleq\{\ETA,\rr,z\}$, $\rr\triangleq\{r_k\}, \forall k$, and $z$ are additional variables. 
		Using the same approach to deal with constraint \eqref{ratedki-lowerbound-1}, we obtain the concave lower bound  $\widehat{R}_{k} (\ETA)$ of $R_{k}(\ETA)$ as $\widehat{R}_{k} \triangleq \frac{\tau_c - \tau_{d,p}}{\tau_c\log 2} B \big[ \log\big(1+\frac{\Upsilon^{(n)}}{\Phi^{(n)}}\big) + \frac{2\Upsilon^{(n)}}{(\Upsilon^{(n)}+\Phi^{(n)})} 
		- \frac{(\Upsilon^{(n)})^2}{(\Upsilon^{(n)}+\Phi^{(n)})\Upsilon}
		- \frac{\Upsilon^{(n)}\Phi}{(\Upsilon^{(n)}+\Phi^{(n)})\Phi^{(n)}}\big]$,
		where $\Upsilon(\eta_k) \triangleq (M-K)\rho \sigma_k^2\eta_{k}$, $\Phi(\ETA) \triangleq \rho (\beta_k - \sigma_k^2) \sum_{\ell\in\K} \eta_{\ell} +1$. Then, constraints \eqref{Rk:lb} can be approximated by the following convex constraint
		\begin{align}
			\label{ratedk-lowerbound-1-a}
			& r_{k} \leq \widehat{R}_{k}(\ETA), \forall k.
		\end{align}
		
		Now, at iteration $(n+1)$, for a given point $\y^{(n)}$, problem \eqref{Pmain:conven:epi} can be approximated by the following convex problem:
		\begin{align}\label{Pmain:conven:epi:cvx}
			\underset{\y\in\widetilde{\HHH}}{\min} \,\,
			& z,
		\end{align}
		where $\widetilde{\HHH}\triangleq\!\{
		\eqref{powerdupperbound:asyn:syn}, \eqref{powerlowerbound:as}, \eqref{z:lb}, \eqref{ratedk-lowerbound-1-a}
		\}$ is a convex feasible set. In Alg.~\ref{alg:conven}, we outline the main steps to solve problem \eqref{Pmain:conven:epi}.
		Let $\HHH\triangleq \{\eqref{powerdupperbound:asyn:syn}, \eqref{powerlowerbound:as}, \eqref{z:lb}, \eqref{Rk:lb}\}$ be the feasible set of problem \eqref{Pmain:conven:epi}. 
		Starting from a random point $\y\in\HHH$, we solve \eqref{Pmain:conven:epi:cvx} to obtain its optimal solution $\y^*$, and use $\y^*$ as an initial point in the next iteration. The algorithm terminates when an accuracy level of $\varepsilon$ is reached. Since $\widetilde{\HHH}$ satisfies Slater's constraint qualification condition, 
		Alg.~\ref{alg:conven} converges to a Karush--Kuhn--Tucker solution of \eqref{Pmain:conven:epi} (hence \eqref{Pmain:conven}) \cite[Theorem 1]{Marks78OR}.
		
		\begin{algorithm}[!t]
			\caption{Solving problem \eqref{Pmain:conven:epi}}
			\begin{algorithmic}[1]\label{alg:conven}
				\STATE \textbf{Initialize}: Set $n\!=\!0$ and choose a random point $\y^{(0)}\!\in\!\HHH$.
				\REPEAT
				\STATE Update $n=n+1$
				\STATE Solve \eqref{Pmain:conven:epi:cvx} to obtain its optimal solution $\y^*$
				\STATE Update $\y^{(n)}=\y^*$
				\UNTIL{convergence}
			\end{algorithmic}
			\vspace{+0mm}
		\end{algorithm}

		
		\vspace{-0mm}
		\subsubsection{Conventional Non-Data~Size-Aware Scheme}
		This scheme does not take into account the size of the UE data. The completion time is reduced by improving the rates of all UEs. To this end, the scheme aims to maximize the lowest rate of all UEs, which leads to the following problem
		\begin{subequations}\label{Pmain:conven2}
			\begin{align}
				\!\!\!\!\!\underset{\ETA}{\max} \,\,
				& 
				\min_{k} {R_{k}(\ETA)}
				\\
				\!\!\!\!\!\mathrm{s.t.}\,\,
				\nonumber
				& \eqref{powerdupperbound:asyn:syn}, \eqref{powerlowerbound:as}.
			\end{align}
		\end{subequations}
		The optimal solution to problem \eqref{Pmain:conven2} can be written in closed-form as $\eta_k = \frac{1+\rho(\beta_k-\sigma_k^2)} {\rho\sigma_k^2 \big(\frac{1}{\rho}\sum_{\ell\in\K} \frac{1}{\sigma_\ell^2} + \sum_{\ell\in\K} \frac{\beta_\ell-\sigma_\ell^2}{\sigma_\ell^2}\big)}$ \cite[Tab. 5.4]{ngo16}. 
		
		\vspace{-0mm}
		\subsubsection{Heuristic Small-Scale-Fading-Based  Scheme}
		In each small-scale coherence block, greedy user scheduling and power allocation to (approximately) maximize the lowest rate are performed.
		Specifically, if the data queue of a UE becomes zero, this UE will be no longer scheduled in the subsequent small-scale coherence block. This way, the remaining UEs at later small-scale coherence blocks will have more power and higher transmission rate, which eventually contributes to reducing the longest completion time. The optimal power control for the max-min rate problem is $\eta_k = 1/c_k/(\sum_{k\in\K}(1/c_k))$, where $c_k = \rho |\g_k^T \uu_k|^2$.

		\vspace{-4mm}
		\section{Numerical Results}
		\vspace{-2mm}
		We consider a square-shaped cell of size $D\times D$, where $D=0.25$ km. The BS is at the center, and the UEs are randomly located. 
		We set $\tau_c\!=\!200$ samples.
		The large-scale fading coefficients, $\beta_{k}$, are modeled as in \cite{3gpp10}, $\beta_k[\text{dB}] = - 148.1  - 37.6 \log_{10}\big(\tfrac{d_k}{1\,\,\text{km}}\big) + z_k$,
		where $d_k \geq 35$ m is the distance between UE $k$ and the BS, and $z_k$ represents shadow fading, which has zero mean and $7$ dB standard deviation. We take the bandwidth to $B = 100$ MHz and the noise power to $\sigma_0^2\!=\!-92$ dBm. Let $\tilde{\rho}\!=\!1$ W and $\tilde{\rho}_p\!=\!0.1$ W be the maximum transmit power of the BS and uplink pilot sequences, respectively. The maximum transmit powers $\rho$ and $\rho_p$ are normalized by the noise power. The sizes of the UE data are taken to monotonically increase from UE $1$ to UE $K$ with a step $\Delta$, i.e., $S_{k+1} = S_k + \Delta$. Here, $S_1 = 0.125$ MB and $\Delta = 0.5$ MB. We set $T_c=1$ms and $\widetilde{T}_c=10$s. 
		
		\begin{figure}[t!]
			\centering
			{\includegraphics[width=0.4\textwidth]{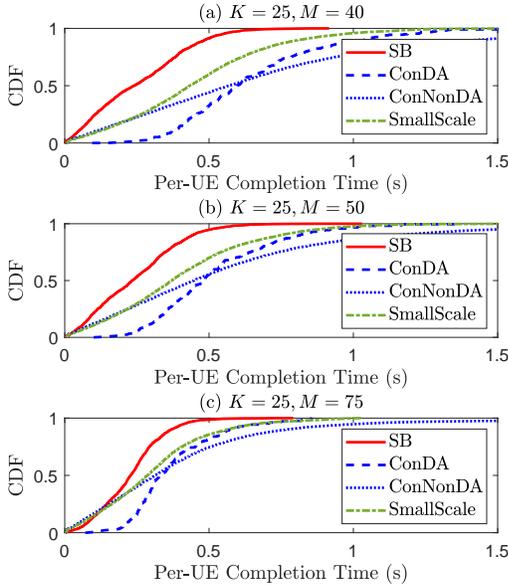}}
			\vspace{-9mm}
			\caption{Comparison of the proposed scheme with baseline schemes.}
			\label{Fig:time}
			\vspace{-0mm}
		\end{figure}
		
		Fig.~\ref{Fig:time} compares the completion time per UE of the proposed session-based (\textbf{SB}) scheme to those of the conventional data size-aware (\textbf{ConDA}), non-data size-aware (\textbf{ConNoDA}) schemes, and heuristic scheme on the small-scale fading time scale (\textbf{SmallScale}). The results in Fig.~\ref{Fig:time} are obtained using $200$ channels. The maximum completion time of UEs of \textbf{ConNoDA} is smaller than $\widetilde{T}_c$ and the minimum is larger than $T_c$. 
		As seen, in terms of $90\%$-likely performance, the session-based scheme significantly outperforms \textbf{ConDA}, \textbf{ConNoDA}, and \textbf{SmallScale} when  $M/K$ is small. Specifically, 
		the $90\%$-likely completion time per UE of \textbf{SB} with $K=25, M=40$ is $0.48$s, which is $2-3$ times smaller than those of \textbf{SmallScale}, \textbf{ConDA} and \textbf{ConNoDA}.
		For the case of $(K=25,M=75)$ where $M/K$ is already large, the inter-user interference is relatively small, and hence, the improvement by our session-based scheme becomes smaller. The baseline schemes can outperform   the proposed scheme in some cases but only in terms of less-than-$25\%$-likely performance. Fig.~\ref{Fig:time} also shows the advantage of joint optimization of user assignment, time, and rates over the small-scale allocation approach.

		\vspace{-3mm}
		\section{Conclusion}
		\vspace{-1mm}
		\label{sec:con}
		In this work, we have proposed a session-based scheme for the massive MIMO downlink, where the BS knows, a priori, the amount of data sent to the UEs. We formulated an optimization problem for assigning UEs to sessions and allocating power to minimize the completion time of the UEs. Utilizing successive convex approximation techniques, we proposed a novel algorithm to solve the formulated problem. Numerical results showed that our session-based scheme can significantly reduce the completion time compared with conventional schemes. 
		\vspace{-0mm}

		\vspace{-4mm}
		\appendix
		\vspace{-1mm}
		The proof basically follows the arguments in \cite{vu18TCOM} with some modifications for our setting. Denote by $\LL_\lambda$ the optimal value of problem \eqref{Pmain:epi:relax} corresponding to $\lambda$. Also, denote by $q^*$ the optimal value of problem \eqref{Pmain:epi}. Then $q^* < + \infty$ due to the compactness of $\FF$. By a duality gap between the optimal values of problem \eqref{Pmain:epi} and its dual problem, 
		\begin{align}
			\nonumber
			\sup_{\lambda\geq 0} \LL_\lambda
			&= \sup_{\lambda\geq 0} \min_{\widetilde{\x} \in \widehat{\FF}} \LL(\aaa, \tilde{\rr}, \tilde{\ttt}, \Ss)
			\leq q^* = \min_{\widetilde{\x} \in \widehat{\FF}} \max_{\lambda \geq 0} \LL(\aaa, \tilde{\rr}, \tilde{\ttt}, \Ss).
		\end{align}
		It follows that, for all $\lambda\geq 0$,
		\begin{align}
			\label{Estar}
			\LL_\lambda \leq q^* < +\infty. 
		\end{align}
		
		For each $\lambda\geq 0$, let $V_{1,\lambda} \triangleq \sum_{k\in\K} \sum_{i\in\K}  ((\tilde{r}_{k,i})_{\lambda} (\tilde{t}_{k,i})_{\lambda} - (S_{k,i})_{\lambda})$ and $V_{2,\lambda} \triangleq \sum_{k\in\NN} \sum_{i\in\K} ((a_{k,i})_{\lambda}-(a_{k,i})_{\lambda}^2))$ be the values of $V_1$ and $V_2$ at the optimal solution $(\aaa_{\lambda}, \tilde{\rr}_{\lambda}, \tilde{\ttt}_{\lambda}, \Ss_{\lambda})$ of \eqref{Pmain:epi:relax} corresponding to $\lambda$. We see from \eqref{ratedki-lowerbound-1}--\eqref{sametime:sessisionS1-1-a} that $V_{1,\lambda} \geq 0$ and from \eqref{arelax} that $V_{2,\lambda} \geq 0$. Set $V_{\lambda} \triangleq \gamma_1 V_{1,\lambda} + \gamma_2 V_{2,\lambda}$ and let $q_{\lambda}$ be the value of $q$ corresponding to $\lambda$. Next, let $0 \leq \lambda_1 < \lambda_2$. By the definition of $\LL_{\lambda_1}$ and $\LL_{\lambda_2}$,
		\begin{align}
			\label{Elambda1}
			&\LL_{\lambda_1} = q_{\lambda_1} + \lambda_1 V_{\lambda_1} \leq  q_{\lambda_2} + \lambda_1 V_{\lambda_2},
			\\
			\label{Elambda2}
			&\LL_{\lambda_2} = q_{\lambda_2} + \lambda_2 V_{\lambda_2} \leq  q_{\lambda_1} + \lambda_2 V_{\lambda_1},
		\end{align}
		from which we have $\lambda_1 V_{\lambda_1} + \lambda_2 V_{\lambda_2} \leq \lambda_1 V_{\lambda_2} + \lambda_2 V_{\lambda_1}$, and so $V_{\lambda_2} \leq V_{\lambda_1}$. This means $V_{\lambda}$ is decreasing as $\lambda$ is increasing. Since $V_{\lambda} = \gamma_1 V_{1,\lambda} + \gamma_2 V_{2,\lambda}\geq 0$ for all $\lambda \geq 0$, we obtain that $V_{\lambda} \rightarrow V^* \geq 0 \text{~as~} \lambda \rightarrow +\infty$.
		From \eqref{Elambda1} and \eqref{Elambda2}, we also have that $\lambda_2 q_{\lambda_1} + \lambda_1 q_{\lambda_2} \leq \lambda_2 q_{\lambda_2} + \lambda_1 q_{\lambda_1}$,
		which yields $q_{\lambda_2} \geq q_{\lambda_1}$. Therefore, as $\lambda \rightarrow +\infty$, $q_{\lambda}$ is increasing and hence bounded from below. 
		Now, if $V^* > 0$, then $\LL_\lambda = q_{\lambda} + \lambda V_{\lambda} \rightarrow + \infty$ as $\lambda \rightarrow + \infty$, which contradicts \eqref{Estar}. Thus, we must have $V^* = 0$, that is, $V_{\lambda}\rightarrow 0$ as $\lambda \rightarrow + \infty$, 
		which implies that $V_{1,\lambda} \rightarrow 0$ and $V_{2,\lambda} \rightarrow 0$ as $\lambda \rightarrow +\infty$.
		
		Finally, let $\widetilde{\x}_{\lambda}$ be the value of $\widetilde{\x}$ corresponding to $\lambda$. Then $\widetilde{\x}_{\lambda}\in \widehat{\FF}$. Since $\widehat{\FF}$ is bounded, there exists a cluster point $\widetilde{\x}_*$ of $\{\widetilde{\x}_{\lambda}\}_{\lambda}$ as $\lambda \rightarrow +\infty$. We assume, without loss of generality, that $\widetilde{\x}_{\lambda} \rightarrow \widetilde{\x}_*$. Then, $\aaa_{\lambda} \rightarrow \aaa_{*}$, $\tilde{\rr}_{\lambda} \rightarrow \tilde{\rr}_{*}$,  $\Ss_{\lambda} \rightarrow \Ss_{*}$, and $\tilde{\ttt}_{\lambda} \rightarrow \tilde{\ttt}_{*}$. It follows that $V_{1,\lambda} \rightarrow (V_{1})_* \!\triangleq\! \sum_{k\in\K} \sum_{i\in\K} ((\tilde{r}_{k,i})_{*} (\tilde{t}_{k,i})_{*} - (S_{k,i})_{*})$, 
		$V_{2,\lambda} \rightarrow (V_{2})_* \!\triangleq\! \sum_{k\in\NN} \sum_{i\in\K} ((a_{k,i})_{*}-(a_{k,i})_{*}^2))$, 
		$V_{\lambda} \rightarrow V_* \triangleq \gamma_1 (V_1)_* + \gamma_2 (V_2)_*$, and $q_{\lambda} \rightarrow q_{*}$. 
		As shown above, $(V_{1})_* = 0$, $(V_{2})_* = 0$, and $V_* = 0$. Therefore, $(\tilde{\rr}_{*}, \tilde{\ttt}_{*}, \Ss_{*})$ and $(\aaa_{*})$ satisfy \eqref{V1} and \eqref{V2}, respectively.
		This together with $\widetilde{\x}_{*} \in \widehat{\FF}$ implies that $\widetilde{\x}_{*} \in \FF$, and so $\widetilde{\x}_{*}$ is a feasible point of \eqref{Pmain:epi}. As such, $q_* \geq q^*$. Next, the definition of $\LL_\lambda$ implies that, for all $\lambda \geq 0$, $\sup_{\lambda\geq 0} \LL_\lambda \geq \LL_\lambda = q_{\lambda} + \lambda V_{\lambda} \geq q_{\lambda}$.
		By letting $\lambda \rightarrow +\infty$, $\sup_{\lambda\geq 0} \LL_\lambda \geq q_* \geq q^*$.
		Combining with \eqref{Estar}, yields $\sup_{\lambda\geq 0} \widetilde{\LL}(\lambda) = q_* = q^*$. We conclude that \eqref{Strong:Dualitly:hold} holds and that $\widetilde{\x}_{*}$ is an optimal solution of \eqref{Pmain:epi}, which completes the proof.  
		


		\ifCLASSOPTIONcaptionsoff
		\newpage
		\fi

		
		\vspace{-3mm}
		\begin{spacing}{1}
			\bibliographystyle{IEEEtran}
			\vspace{-0mm}
			\bibliography{IEEEabrv,newidea2021}
		\end{spacing}
		
	\end{document}